\documentclass[acus]{JAC2003}

\usepackage[final]{graphicx}
\usepackage{booktabs}
\usepackage[pdftex,final]{hyperref}
\graphicspath{{figures/}}
\usepackage{siunitx}
\DeclareSIUnit{\dBm}{dBm}
\usepackage{xspace}

\usepackage{xcolor}

\begin{document}
\title{SCANNING WIRE BEAM POSITION MONITOR FOR ALIGNMENT OF A HIGH BRIGHTNESS INVERSE-COMPTON X-RAY SOURCE\thanks{ Work supported by the US Department of Homeland Security DNDO ARI program GRANT NO. 2010-DN-077-ARI045-02 }}

\author{M.~R. Hadmack\thanks{hadmack@hawaii.edu} and E.~B. Szarmes\\
University of Hawai`i Free-Electron Laser Laboratory, Honolulu, HI 96822, USA}

\maketitle

\begin{abstract}
The Free-Electron Laser Laboratory at the University of Hawai`i has constructed and tested a scanning wire beam position monitor to aid the alignment and optimization of a high spectral brightness inverse-Compton scattering x-ray source.  
X-rays are produced by colliding the \SI{40}{\MeV} electron beam from a pulsed S-band linac with infrared laser pulses from a mode-locked free-electron laser driven by the same electron beam.  
The electron and laser beams are focused to \SI{60}{\um} diameters at the interaction point to achieve high scattering efficiency.  
This wire-scanner allows for high resolution measurements of the size and position of both the laser and electron beams at the interaction point to verify spatial coincidence.  
Time resolved measurements of secondary emission current allow us to monitor the transverse spatial evolution of the e-beam throughout the duration of a \SI{4}{\us} macro-pulse while the laser is simultaneously profiled by pyrometer measurement of the occulted infrared beam.  
Using this apparatus we have demonstrated that the electron and laser beams can be co-aligned with a precision better than \SI{10}{\um} as required to maximize x-ray yield.
\end{abstract}

\section{INTRODUCTION}
A compact high brightness x-ray source is currently under development at the University of Hawai`i Free-Electron Laser Laboratory, based on inverse-Compton scattering of \SI{40}{\MeV} electron bunches with synchronous laser pulses from an infrared free-electron laser (FEL)\cite{hadmack2012phd,madey2013oce}.
One of the more challenging aspects of realizing a Compton backscatter x-ray source is co-alignment of the electron and laser beams.  
With high intensities, and spot sizes as small as \SI{30}{\um}, it is not possible to align these beams without special diagnostic tools. 
The resolution of available beam position monitors (BPMs) and optical transition radiation (OTR) screens is limited to about \SI{100}{\um} by the sampling electronics and video cameras used.

Wire scanners are commonly employed on accelerator beam-lines as alignment aides.  
The ``flying wire'' type scanners, such as those used at CERN, are too large for use in the space allocated on the Mk~V beam-line at UH and are incompatible with the bunch structure of our accelerator.  
The x-ray interaction point is shared by two other insertable diagnostic devices in a crowded vacuum chamber, also housing the x-ray interaction point laser optics.  
The wire scanner described here is based on the designs used at NBS-LANL\cite{cutler1987performance} and the SLC\cite{ross1991wire} and adapted to the constraints of our beamline configuration.
This system also includes the capability to resolve the temporal evolution of the electron beam profile over the macropulse duration (approximately \SI{4}{\us}).

\section{HARDWARE}
\begin{figure}[htbp]
	\centering
	\includegraphics[width=75mm]{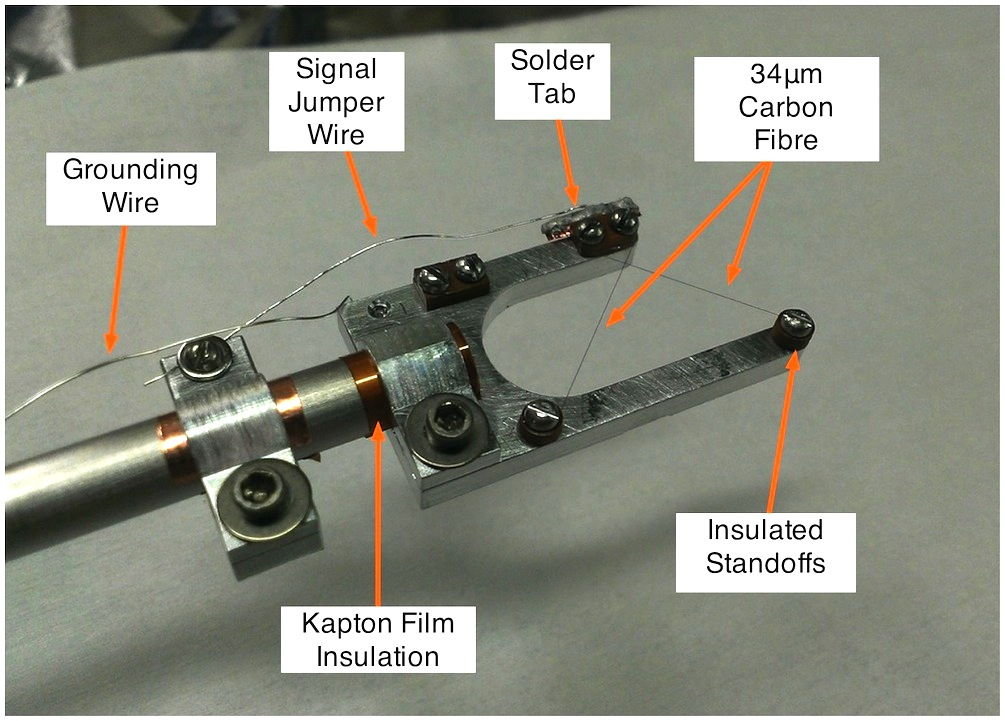}
	\caption{The wire scanner fork electrically isolates the carbon fiber from the grounded fork.  Fibers are soldered to the signal lead and clamped at the other end.}
	\label{fig:swbpm_fork_photo}
\end{figure}

The wire scanner head shown in Fig.~\ref{fig:swbpm_fork_photo} consists of two \SI{34}{\um}-diameter graphite fibers stretched across the \SI{12.3}{\mm} gap in an aluminum fork.  
The wires are oriented such that when the scanner insertion axis is inclined \ang{45} above the beam plane, the two wires are oriented horizontally and vertically.  
In this way a single axis of motion allows the beam to be scanned in both axes.  

\begin{figure}[htbp]
	\centering
	\includegraphics[width=80mm]{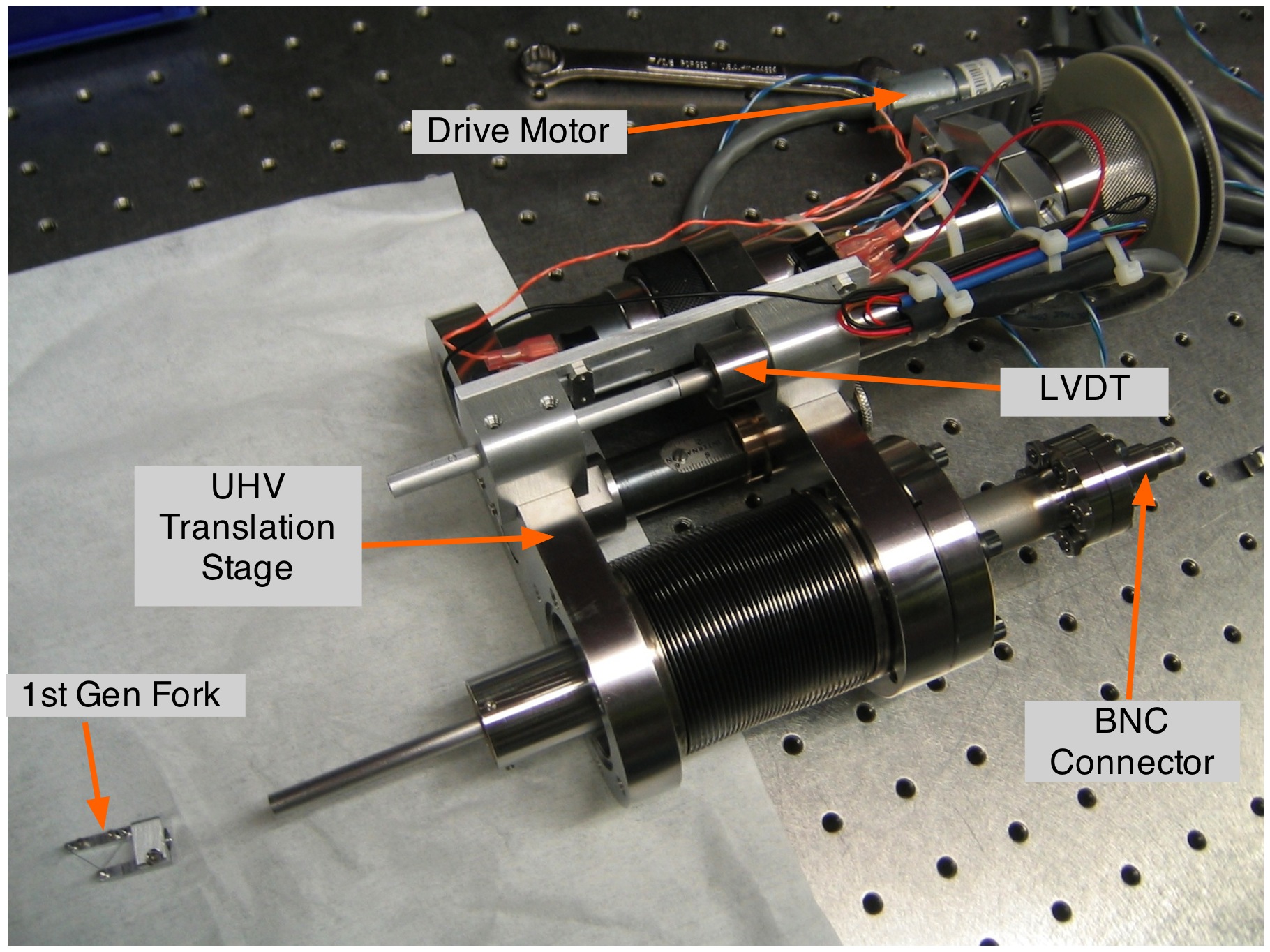}
	\caption{Beam profiler drive assembly with motor and LVDT with an early prototype fork.  The support rod conducts the signal to the vacuum feedthrough on the far end.}
	\label{fig:swbpm_assembly_photo}
\end{figure}

The secondary emission current from the wire is conducted via the scanner shaft to a vacuum feedthrough on the assembly shown in Fig.~\ref{fig:swbpm_assembly_photo}.
The fork itself is grounded to avoid charge accumulation from the beam halo.

\begin{figure}[htbp]
	\centering
	\includegraphics[width=85mm]{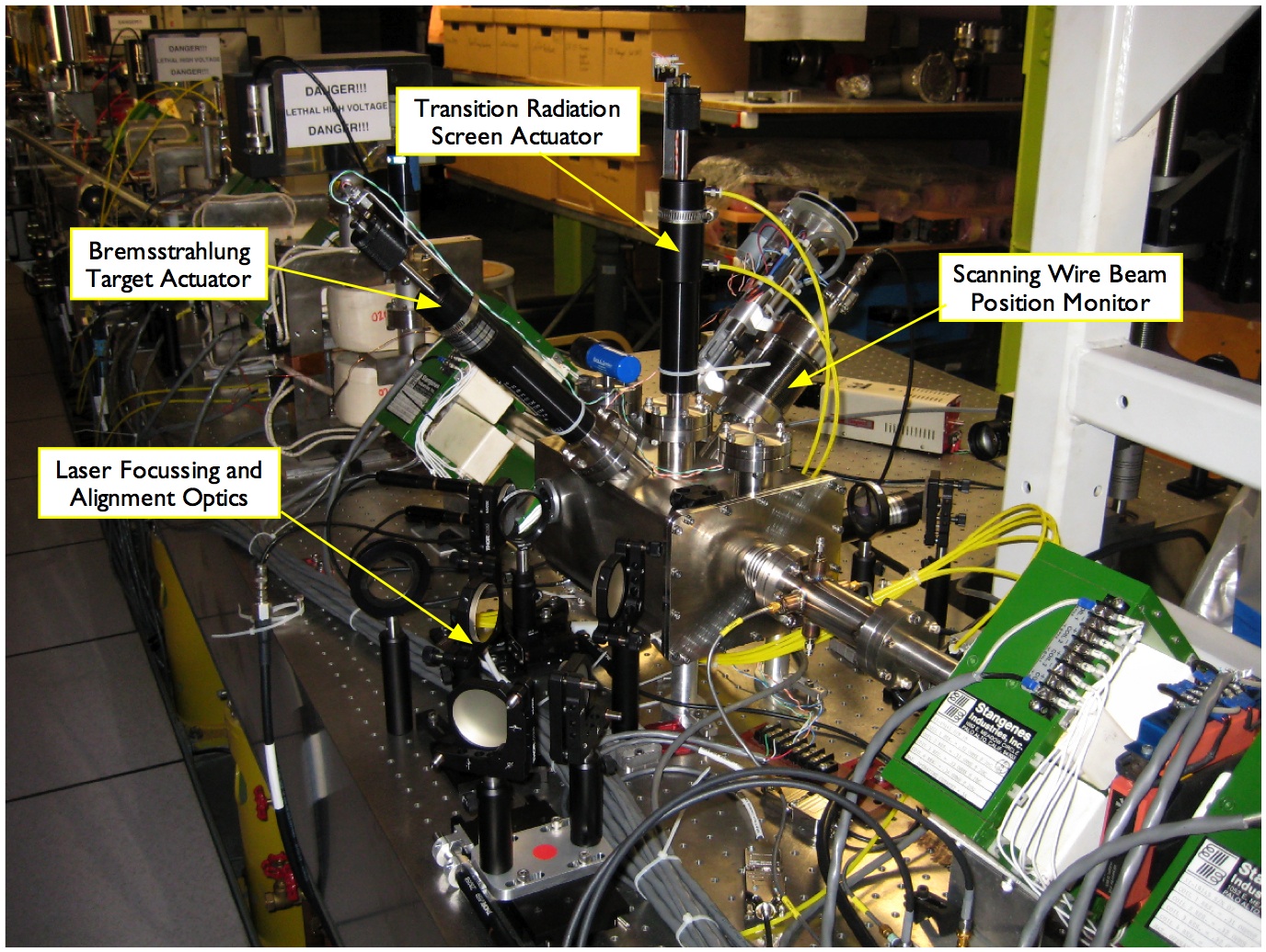}
	\caption{The wire scanner installed with other diagnostic devices in the x-ray scattering chamber.}
	\label{fig:chamberinstalled}
\end{figure}

Most wire scanners in operation today utilize bremsstrahlung radiation detectors to measure beam interception of the wire. 
However, on a linear machine it is more difficult to position a PIN diode detector close to the source without substantial background radiation.
Moving the detector to a suitably shielded location results in a large bremsstrahlung beam diameter, making efficient detection difficult and introducing errors due to diffraction.

Figure~\ref{fig:chamberinstalled} shows the assembly integrated in the x-ray scattering chamber installed at the interaction point.
The wire scanner assembly consists of a precision vacuum translation stage, a linear-variable-differential-transformer (LVDT) position sensor, and a DC motor drive.  

The motor speed is controlled by software to achieve high resolution at the \SI{5}{\Hz} beam repetition rate.
Position is measured with an LVDT attached to the translation stage; its resolution is limited by 12~bit readout electronics to \SI{7.2}{\um} steps over a \SI{14}{\mm} range.
The LVDT read-back is calibrated against the actual translation stage motion with calipers.

Wire scans are typically performed with the actuator speed set to \SI{100}{\um\per\second} so that the position changes by twice the LVDT limiting step size each macropulse event, thus ensuring monotonic position data. 
A full \SI{14}{\mm} scan using both the horizontal and vertical wire takes approximately \SI{140}{\second}.  
For each accelerator macropulse trigger three quantities are measured: the current from the wire, the laser pulse transmission, and the position.  
The intercepted electron beam current is inferred by the current resulting from secondary electrons ejected from the wire.  
The wire current signal is terminated in \SI{50}{\ohm} and sampled with a \SI{300}{\MHz} digital oscilloscope.  

A pyroelectric detector viewing the transmitted beam measures the occlusion of the laser beam by the wire.
The pyrometer's response time is considerably slower than that of the wire current.
The pulse peak voltage is sampled with a boxcar integrator and digitized with 12~bit precision.

The data acquisition software is implemented in Python with a graphical user interface (GUI) built using wxPython.  
The software acquisition is triggered using control lines on an RS232 serial port to monitor the accelerator's TTL trigger.  
When a trigger event is detected, data is read from the oscilloscope and boxcar integrator, both of which are synchronously triggered.  
Data is also acquired asynchronously from the LVDT controller via a serial connection.

\begin{figure}[htbp]
	\centering
		\includegraphics[width=65mm]{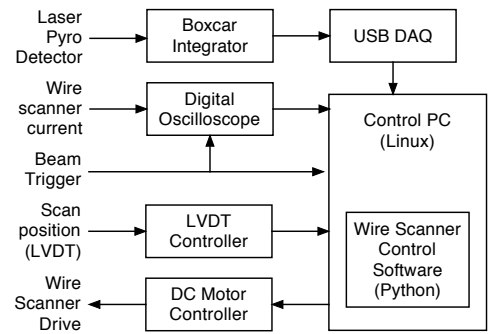}
	\caption{A PC acquires beam current, laser intensity, and scan position data and controls the drive motor.}
	\label{fig:swbpm_blockdiagram}
\end{figure}

\begin{figure}[htbp]
	\centering
	\includegraphics[width=80mm]{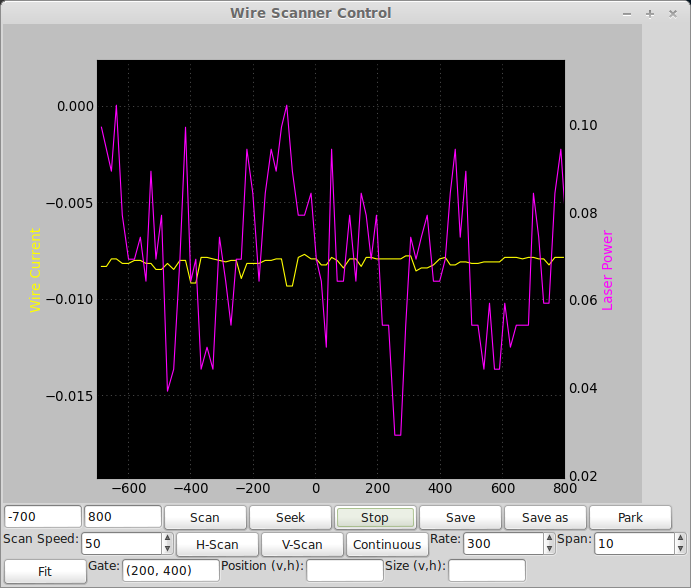}
	\caption{The GUI displays the electron beam (yellow) and laser (purple) beam profiles in real time during a scan.  Figure shows background data, not actual scan results.}
	\label{fig:swbpm_gui}
\end{figure}

Figure~\ref{fig:swbpm_blockdiagram} illustrates the data acquisition system.  
The GUI shown in Fig.~\ref{fig:swbpm_gui} provides the operator with a live stripchart of both the laser and current measurements throughout a scan.  
The GUI allows for configuration and control of automated scans and data storage.  
Data is stored in a custom binary format and includes full oscilloscope waveforms for every position step.

\begin{figure}[htbp]
	\centering
	\includegraphics[width=85mm]{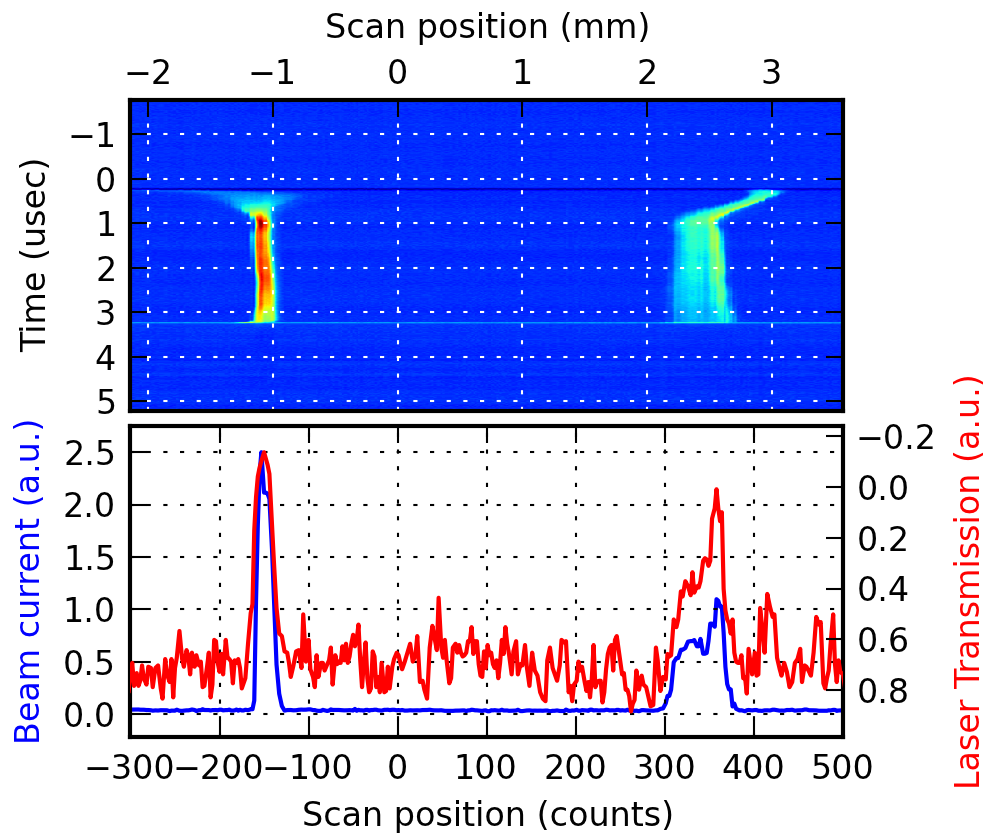}
	\caption{Secondary emission current as a function of position and time within the macropulse.  The lower plot compares the integrated current from \SIrange{2}{3}{\us} with the laser signal (axis inverted).  This scan includes both the horizontal (right) and vertical (left) profiles.}
	\label{fig:scan1}
\end{figure}

The GUI stripchart is only used as a rough guide for scan operation while detailed data analysis is performed using an offline tool, also implemented in Python.  
Figure~\ref{fig:scan1} shows a sample wire scan analysis.  
The graphic in the upper part of the figure is a representation of the evolution of the beam current spatial distribution over a \SI{4}{\us} macropulse.  
The vertical columns in the image are individual oscilloscope waveforms for each position along the horizontal axis; interpolation is applied to account for non-uniformly spaced positions.
The lower plot shows the transmitted laser pulse energy compared to the wire current integrated over a particular duration of interest within the pulse.
The integration region is typically chosen to overlap the laser pulse in the last \si{\us} of the macropulse.

\section{RESULTS}
Preliminary experiments have been conducted to measure the sizes, positions, and stability of the laser and electron beams.
The wire scanner was initially commissioned with a \SI{200}{\um} tungsten wire.
This wire was operated successfully for several months with large beam diameters and limited resolution; however, a microfocused (sub \SI{100}{\um} diameter) beam quickly severed the wire.
Next, a \SI{25}{\um} tungsten wire was chosen to reduce the absorption volume and  to enhance the spatial resolution of the scans by a factor of eight.
Again, however, a microfocused beam destroyed the wire on the first pass.
Ultimately, \SI{34}{\um}-diameter carbon monofilament from Specialty Materials, Inc. has proven robust enough to endure the highly focused \SI{150}{\mA} electron beam and several \si{\milli\joule} of infrared laser exposure.
Since the carbon filament could not be wrapped in the same manner as the more flexible tungsten, it was necessary to modify the wire scanner fork.
Figure~\ref{fig:swbpm_fork_photo} illustrates how the carbon fibers are clamped on one end between Vespel plastic discs while the other ends are soldered to a copper tab attached to the signal lead.

The scan data in Fig.~\ref{fig:scan1} shows an electron beam focused to $w_x,w_y = \SI{350}{\um},\SI{115}{\um}$, where the left feature is the evolution of the vertical beam profile and the right represents the horizontal.
The total charge intercepted on each wire is the same, so the area under the beam profile curves is constant, resulting in lower peak signal for the horizontal scan.
In this scan the laser is well aligned to the electron beam resulting in suppression of laser operation while the beam is intercepted by the wire.

\begin{figure}[htbp]
	\centering
	%
	\includegraphics[width=85mm]{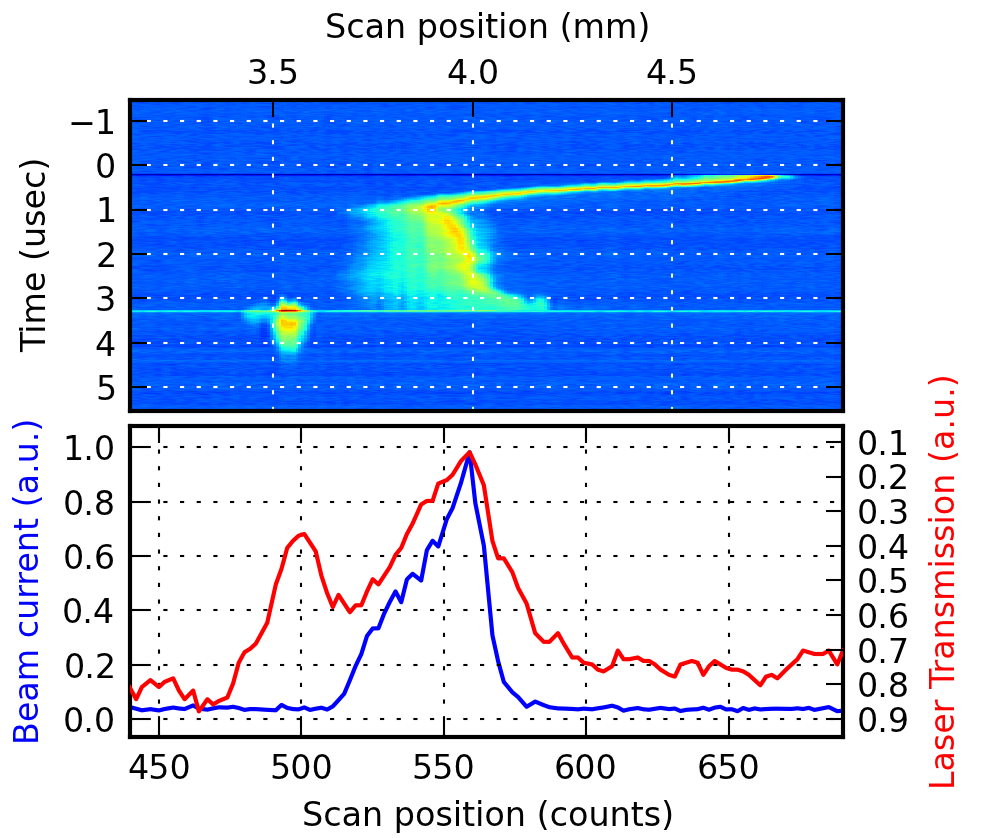}
	\caption{Horizontal axis wire scan showing the laser displaced from the e-beam position.  
	Interception of the e-beam by the wire inhibits lasing and results in a second co-aligned peak.}
	\label{fig:scan2}
\end{figure}

\begin{figure*}[htbp]
    \centering
    \includegraphics*[width=160mm]{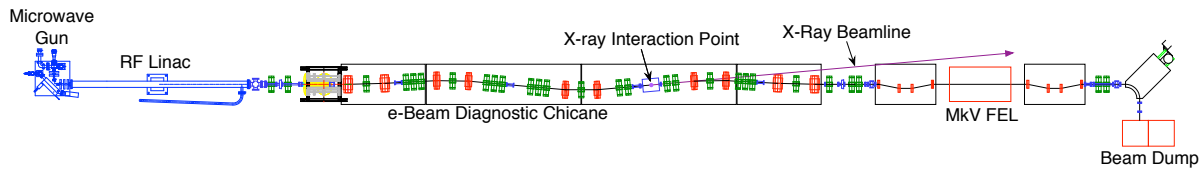}
    \caption{The Mark~V FEL beamline at the University of Hawai`i}
    \label{fig:markv}
\end{figure*}

Figure~\ref{fig:scan2} shows an example where the laser is misaligned from the electron beam.
This horizontal scan (vertical wire only) gives an electron beam width of \SI{175}{\um} and a laser beam width of \SI{143}{\um} with a beam separation of \SI{419}{\um}, correctable with a motorized mirror.
The laser was intentionally defocused at the interaction point for this scan to preclude wire damage during these early experiments.
It is interesting to note that while the laser size and position are measurable by transmission to the pyroelectric detector (red trace), a laser signal also appears distinctly in the wire current.
We believe that this laser induced wire current is the result of thermionic emission from the wire due to heating from laser exposure.
This hypothesis is supported by the observation that the laser induced current extends several hundred \si{\ns} beyond both the end of the electron beam and laser macropulses while in fact the laser beam exposure of the wire begins a microsecond earlier and is therefore presumed to be a thermal artifact.
The photon energy of the \SI{3000}{\nm} laser is not sufficient to generate photoelectrons.
This feature provides a useful way to measure both laser and electron beams with a single sensor signal.

Figure~\ref{fig:scan2} also illustrates the time resolved nature of this wire scanner system.  
The trajectory in this image indicates that the beam's horizontal position slews nearly a \si{\mm} over the \SI{4}{\us} macropulse.  
The large slew in the first microsecond is an inevitable consequence of beam-loading in the linear accelerator and is typically ignored for experimental purposes.
The remainder of the macropulse, however, also shows a position slew that is the consequence energy slew in the beam.
The diagnostic chicane shown in Fig.~\ref{fig:markv} contains a number dipole bend magnets upstream of the interaction point that produce energy dependent deflections in the beam.
Characterization and mitigation of this energy/position slew in the beam are of critical importance to operation of the free-electron laser and for beam alignment of the inverse-Compton scattering interaction.
A na\"ive integration of the wire current or a transition radiation image would significantly overestimate the instantaneous beam size as well as produce an ambiguity in the centroid position during the time interval of interest for scattering.
Even with a significant transverse evolution in the beam, the instantaneous size and position can be precisely measured.

The ``flying wire'' type beam profilers employed on many large accelerators and storage rings employ a high velocity wire capable of scanning many stored bunches in a single sweep\cite{igarashi2002aa}.
While this technique enables much faster profile acquisition, the position becomes correlated with a particular time within the bunch train.
Thus, these systems are not capable of revealing the temporal structure of the macropulse in the manner described above and can overestimate the beam size.
Typically this is not a problem for a storage ring that is filled with nearly identical bunches.  However, the transient beam loading experienced in a linac with a thermionic gun results in beam evolution that must be considered.

Wire scan repeatability was measured from the analysis of a scan sequence with the same e-beam configuration.  
The beam centroid can be measured with an uncertainty of $\sigma_{x,y} = \SI{9}{\um}$ and a beam width uncertainty of $\sigma_w = \SI{4}{\um}$.
Scans are always performed in the same direction to eliminate hysteresis due to a \SI{30}{\um} backlash in the translator lead-screw.

\section{CONCLUSION}
A scanning wire beam position monitor has been successfully constructed and operated at the University of Hawaii Free-Electron Laser Laboratory.
This custom design satisfies the tight space restrictions imposed by the need to share access to the interaction point of the inverse-Compton x-ray source with other diagnostic devices and laser optics.
The use of a commercially available linear vacuum translator significantly reduces the engineering time and cost of the system.
\SI{34}{\um} carbon fiber has been selected as a suitable material for scans of a sub-\SI{100}{\um} microfocused electron beam operated at \SI{40}{\MeV} with \SI{150}{\mA} average current over \SI{4}{\us} macropulses.

Further studies of the carbon filament damage threshold for both electron beam and laser exposure are necessary for effective optimization of the inverse-Compton scattering interaction point focus.
The laser can easily be attenuated to avoid wire damage once this is known.  However, the electron beam current cannot be varied and the damage threshold will impose a limit to the average current density allowed on the wire. 
In principle, the \SI{7}{\um} width resolution is sufficient to achieve the \SI{30}{\um} focal spot specification of the UH x-ray source.
Alignment of the beams can be verified with a scan repeatability of better than \SI{10}{\um} when the \SI{30}{\um} hysteresis is accounted for.

Combining time-resolved wire scans with quadrupole magnet scans will give us the capability to perform time-resolved emittance measurements of the e-beam.  This will be a vital capability for our continued efforts to improve extended pulse length thermionic electron gun technology\cite{kowalczyk2013las}.


\section{ACKNOWLEDGMENT}
We acknowledge \href{http://specmaterials.com/}{Specialty Materials, Inc.} for providing the carbon monofilament and John M.~J.~Madey for his advice and operational support on this project.




\end{document}